\documentclass[aps,prb,twocolumn,showpacs,floatfix,groupedaddress,superscriptaddress]{revtex4-2}
\pdfoutput=1
\usepackage[linkcolor=blue,colorlinks=true,breaklinks=true,citecolor=blue,urlcolor=blue]{hyperref}
\usepackage{graphicx,graphics}  
\usepackage{dcolumn}   
\usepackage{braket}
\usepackage[mathscr]{euscript}
\usepackage{amsbsy}
\usepackage{bm} 
\usepackage{amssymb} 
\usepackage[T1]{fontenc}
\usepackage{amsmath}
\usepackage{xcolor}
\usepackage[caption=false,labelformat=simple]{subfig}

\makeatletter
\newcommand{\phantomlabelabovecaption}[2]{
	\protected@write\@auxout{}{
		\string\newlabel{#2}{
			{\number\numexpr\thefigure+1\relax#1}{\thepage}
			{\number\numexpr\thefigure+1\relax#1}{#2}{}
		}
	}
	\hypertarget{#2}{}
}
\makeatother

\begin{document}

\title{Magnetotransport properties of a twisted bilayer graphene in the presence of external electric and magnetic field}
\author{Priyanka Sinha}
\email{sinhapriyanka2012@gmail.com}
\affiliation{Department of Physical Sciences, Indian Institute of Science Education and Research Kolkata\\ Mohanpur-741246, West Bengal, India}
\author{Ayan Mondal}
\email{ayanmondal367@gmail.com}
\affiliation{Department of Physical Sciences, Indian Institute of Science Education and Research Kolkata\\ Mohanpur-741246, West Bengal, India}
\author{Sim\~ao Meneses Jo\~ao}
\affiliation{Department of Materials, Imperial College London\\South Kensington Campus, London SW7 2AZ, United Kingdom}
\author{Bheema Lingam Chittari}
\email{bheemalingam@iiserkol.ac.in}
\affiliation{Department of Physical Sciences, Indian Institute of Science Education and Research Kolkata\\ Mohanpur-741246, West Bengal, India}
\date{\today}
\begin{abstract}
We extensively investigate the electronic and transport properties of a twisted bilayer graphene when subjected to both an external perpendicular electric field and a magnetic field. Using a basic tight-binding model, we show the flat electronic band properties as well as the density of states (DOS), both without and with the applied electric field. In the presence of an electric field, the degeneracy at the Dirac points is lifted where the non-monotonic behavior of the energy gap exists, especially for twist angles below 3$^\circ$. We also study the behavior of the Landau levels (LL) spectra for different twist angles within a very low energy range. These LL spectra get modified under the influence of the external electric field. Moreover, we calculate the dc Hall conductivity ($\sigma_{xy}$) for a very large system using the Kernel Polynomial Method (KPM). Interestingly, $\sigma_{xy}$ makes a transition from a half-integer to an integer quantum Hall effect, \textit{i.e.} the value of $\sigma_{xy}$ shifts from  $\pm 4(n+1/2)~(2e^2/h)$ ($n$ is an integer) to $\pm 2n~(2e^2/h)$ around a small twist angle of $\theta=2.005^\circ$. At this angle, $\sigma_{xy}$ acquires a Hall plateau at zero Fermi energy. However, the behavior of $\sigma_{xy}$ remains unaltered when the system is exposed to the electric field, particularly at the magic angle where the bands in both layers can hybridize and strong interlayer coupling plays a crucial role.
\end{abstract}
\maketitle

\section{Introduction}
After the discovery of graphene \cite{wallace,neto,geim}, the van der Waals (vdW) layered materials \cite{geim2} have been at the forefront of scientific interest in recent years. Among these, twisted bilayer graphene (TBG) \cite{santos,Bistritzer, Andrei, chakov} is one of such heterostructures which stands out due to its remarkable band properties. These properties encompass the coexisting massive and massless Dirac fermions \cite{kim}, second-generation Dirac singularities \cite{mele}, the Hofstadter butterfly spectrum \cite{Bistritzer2}, the coexistence of superlattice Dirac points and van Hove singularities (VHS) \cite{chu} etc. In TBG, two graphene layers are rotated at a specific angle relative to each other, resulting in the formation of a moir\'e pattern. This pattern can be visualized using scanning tunneling microscopy \cite{reina,miller} and exhibits a periodic nature, with larger periods observed at smaller twist angles \cite{hermann}. Moreover, the twist angle impacts the position of the Dirac cones in each layer, causing a displacement in momentum \cite{yan2}.\par
\begin{figure*}[!ht!]
	\begin{center}
		\phantomlabelabovecaption{(a)}{fig:sc_1}
		\phantomlabelabovecaption{(b)}{fig:sc_2}
		\phantomlabelabovecaption{(c)}{fig:sc_3}	
		\includegraphics[width=0.9\textwidth]{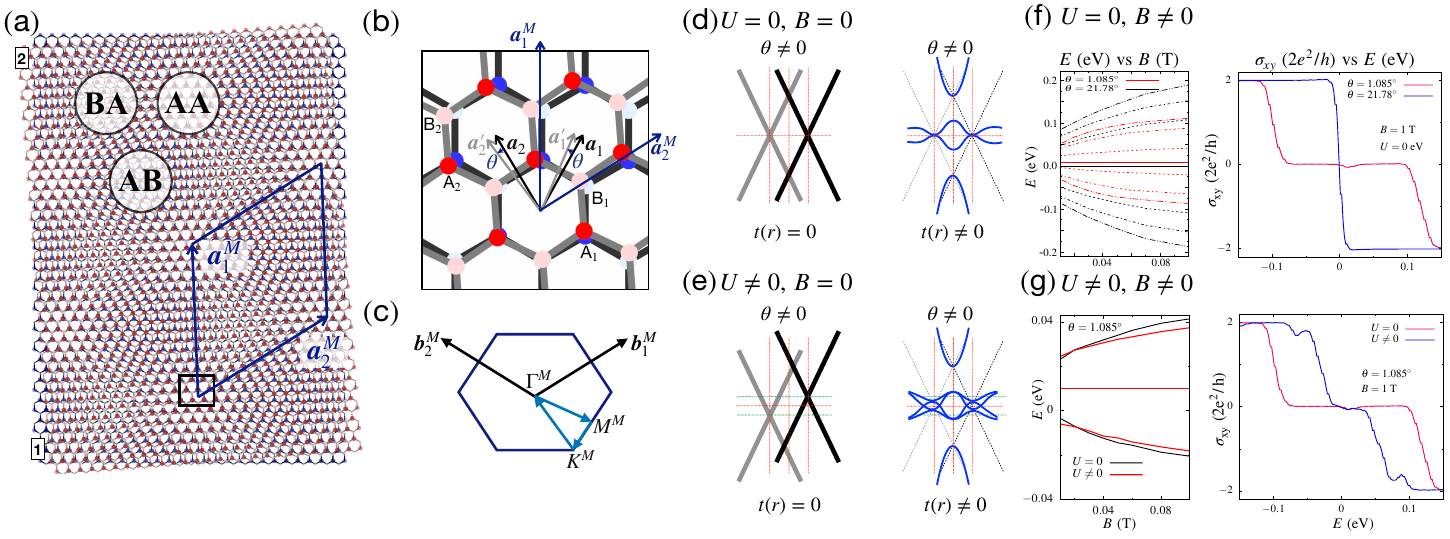}
		\caption{(Color online) (a) A schematic diagram of a TBG lattice is shown. The upper layer (denoted as 2) is rotated with respect to the lower layer (denoted as 1) by a certain angle. The three different stacking regions (referred to as AA, AB, and BA) formed in the superlattice structure are outlined by a black circle. (b) An enlarged view of the rectangular box is shown. The two different sublattices in layer 1 are denoted by $A_{1}$ (blue) and $B_{1}$ (light-blue) and in layer 2 are denoted by $A_{2}$ (red) and $B_{2}$ (light-red). $\boldsymbol{a}_{1}$ and $\boldsymbol{a}_{2}$ are the lattice vectors in layer 1, whereas $\boldsymbol{a}'_{1}$ and $\boldsymbol{a}'_{2}$ are the lattice vectors in layer 2. $\boldsymbol{a}^{M}_{1}$ and $\boldsymbol{a}^{M}_{2}$ are the unit cell lattice vectors for the moir\'e superlattice. $\theta$ corresponds to the rotation angle. (c) The mini Brillouin zone (BZ) of the moir\'e superlattice in reciprocal space is spanned by the vectors $\boldsymbol{b}^{M}_{1}$ and $\boldsymbol{b}^{M}_{2}$. The high-symmetry points ($\Gamma^{M}$, $K^{M}$, $M^{M}$) are given in the Brillouin zone and the arrows indicate the path over the Brillouin zone. Schematics for band structure is shown without ($t(r)=0$) and with ($t(r)\neq 0$) interlayer coupling for (d) $U=0$, $B=0$ and (e) $U\neq 0$, $B=0$. It is shown that for $t(r)=0$, the uncoupled graphene layers doped to opposite charges under the applied electric field. However, for $t(r) \neq 0$, these layers couple with each other and the low energy bands induce gaps at Dirac points in the presence of the applied electric field. Landau levels spectra ($E$ (eV) vs $B$ (T)) and Hall conductivity ($\sigma_{xy}$ (2e$^2/h$) vs $E$ (eV)) are shown for (f) $U=0$, $B \neq 0$ with $\theta=1.085^\circ$ and 21.78$^\circ$, and (g) $U\neq 0$, $B \neq 0$ with $\theta=1.085^\circ$. The Landau levels are sensitive to the twist angle and lead to new quantum hall phases in the system.}
		\label{fig:schematic}
	\end{center}
\end{figure*}
Several methods including growth on the C face of a SiC substrate \cite{hass,sprinkle}, chemical vapor deposition (CVD) growth techniques on metal substrates \cite{li2,lu2,iwasaki,sun}, or mechanical folding of single graphene sheet \cite{carozo} etc. have been used to produce TBG structures. Collective charge carrier behavior arises from the flat regions of the band structure near a saddle point, where multiple electrons share the same energy level being highly localized at the AA sites \cite{trambly}. Experimentally, the localization of electrons and the VHS in twisted graphene layers have been measured by scanning tunneling spectroscopy \cite{reina,luican}. The nearly flat low energy bands in the moir\'e pattern formed by twisted graphene layers, referred to as magic-angle twisted bilayer graphene (MATBG) \cite{Bistritzer}, give rise to a physics of strong correlation \cite{ash,dante}. There are evidences of Mott-like insulating behavior \cite{cao2}, superconductivity \cite{cao1}, anomalous Hall effect \cite{Serlin}, etc. which corroborate the richness of physics it possesses.\par
The low energy electronic structure of TBG possesses massless Dirac fermions, albeit with a diminished Fermi velocity compared to monolayer graphene at small twist angles \cite{santos,shall2,trambly}, which is quite different from that of an AB-stacked bilayer graphene. Both theoretical and experimental studies have demonstrated the alteration of the band gap in a bilayer graphene owing to the different onsite energies between the layers \cite{Ohta,zhang}. There are records of the possibility of controlling the band gap of Bernal bilayer graphene by applying the electric field \cite{kuz,Castro,Mak}. However, contrary to this, an external electric field does not induce a band gap in the band structure of TBG \cite{moon}. Nevertheless, the presence of an interlayer bias gives rise to exotic phenomena, such as additional anisotropic reduction of the Fermi velocity \cite{xian} and the emergence of topologically protected helical modes in the electronic spectrum \cite{San-Jose}. Recent findings also suggest that large angle TBG may offer semiconductor-like behavior with tunable band gaps up to terahertz frequencies when subjected to an electric field \cite{Talkington}. The combination of gate tunability \cite{bhem2, bhem3, bhem4} and twist angle in certain two-dimensional systems has demonstrated the possibility of achieving tunable Mott insulators \cite{Chen}, as well as topological flat bands \cite{Chittari2} etc.\par 
On the other side, the discovery of an unusual quantum Hall effect in monolayer graphene provides evidence for the existence of massless Dirac fermions \cite{novo}, whereas, in the case of bilayer graphene, the quantization rules for the integer quantum Hall effect are quite different confirming the characteristics of massive quasiparticles \cite{McCann}. The Landau level (LL) energies vary as $E_{n} \sim \pm \sqrt{|n|B}$ ($B$ is the magnetic field, $n$ is the LL index) and $E_n \sim \pm \hbar \omega \sqrt{n(n-1)}$ ($\omega=eB/m$ is the cyclotron frequency) for the monolayer and bilayer graphene respectively. However, the role of interlayer coupling becomes crucial in the formation of LL spectra as it influences the behavior of quasiparticles of a pristine TBG \cite{santos, choi}. Strikingly, the massless Dirac fermions still survive in TBG for large twist angles due to the decoupling of twisted layers \cite{hass2}. When exposed to a perpendicular magnetic field, the electronic behavior of TBG remains similar to that of a single layer for twist angles exceeding 20$^\circ$ \cite{luican}. Experimentally, the LL behavior has been addressed using scanning tunneling microscopy and LL scanning tunneling spectroscopy depending on the twist angle \cite{luican}. Furthermore, several studies have been reported on the LL spectrum as well as the Hall quantization of TBG, both through analytical and experimental approaches \cite{gail,moon2,lee,hejazi,choi,senthil}. However, the comprehensive study of the combined effect of electric field and magnetic field in a twisted bilayer graphene has not been addressed so far.\par
In this paper, we investigate the effects of a perpendicular electric field and a magnetic field on the electronic and transport characteristics using the tight-binding model of a TBG system. We predict the first magic angle and study the electronic band structure and DOS characteristics pertaining to the magic angle, as well as for adjacent small twist angles (both below and above the magic angle). Next, we introduce a perpendicular electric field into the system to investigate its influence on the electronic properties. Notably, we observe non-monotonic dispersion of energy levels with varying electric field strength. In the presence of a weak magnetic field, we show the DOS spectrum for the magic angle and also investigate the LL spectra for different twist angles. We also unveil novel effects of the electric field on the LL spectrum for the magic angle. Subsequently, we employ the real-space method to calculate the dc Hall conductivity within the linear response regime, utilizing the Kubo-Bastin formula \cite{kubo1,kubo2}, as implemented in KITE \cite{simao,pires,pires2}. Nevertheless, we reveal the transport behavior in the TBG system with a significant number of unit cells. This underscores the complexities that emerge from modeling of such structures, especially near the magic angle, which creates a high demand for research. \par 
The paper is organized as follows. The geometry and the model Hamiltonian of TBG are introduced in Sec.~(\ref{II}). In Sec.~(\ref{III}), we describe the formalism to compute the transport properties and the DOS in detail. Our results are presented in Sec.~(\ref{IV}). Section~(\ref{A}) and (\ref{B}) include the results for the band structures and the DOS in the absence ($U=0$) and presence ($U \neq 0$) of an external electric field, respectively. The effects of a perpendicular magnetic field ($B \neq 0$) are depicted through the LL spectra shown in Sec.~(\ref{C}). To illustrate the effects of both electric and magnetic fields simultaneously on the LL spectra, we observed the density of states in Sec.~(\ref{D}). Further, the transport properties are investigated by computing the dc Hall conductivities in Sec.~(\ref{E}). Finally, our findings are summarized in Sec.~(\ref{V}).
\section{Model}\label{II}
In this section, we begin by briefly describing the geometrical structure of our system as given in Fig.~\ref{fig:schematic}. Fig.~\ref{fig:sc_1} shows the schematic diagram of a TBG where we have rotated the upper layer (layer 1) of an AA-stacked bilayer graphene (where each atom in one layer aligns directly above or below an atom in the other) in an anticlockwise direction by an angle $\theta$ with respect to the fixed lower layer (layer 2). The two sublattices in the lower layer are labeled as A$_{1}$ (blue) and B$_{1}$ (light-blue), while those in the upper layer are A$_{2}$ (red) and B$_{2}$ (light-red), respectively. The unit cell of a lower layer is spanned by the lattice vectors $\boldsymbol{a}_{1}=(1/2,\sqrt{3}/2)a$ and $\boldsymbol{a}_{2}=(-1/2,\sqrt{3}/2)a$ as shown in Fig.~\ref{fig:sc_2}. Here, $a=2.46$~\AA~being the lattice constant of monolayer graphene and $|\boldsymbol{a}_{1}|=|\boldsymbol{a}_{2}| = a$. Alternatively, $\boldsymbol{a}'_{1}$ and $\boldsymbol{a}'_{2}$ represent the lattice vectors of the rotated upper layer \cite{zhu}. Furthermore, the Brillouin zone of the moir\'e superlattice gets reduced in size with a hexagonal shape when compared to the Brillouin zones of the two separate layers as shown in Fig.~\ref{fig:sc_3}.\par
For a commensurate lattice structure within TBG, the lattice vectors in the unit cell can be expressed as $\boldsymbol{a}^{(M)}_{1}= n\boldsymbol{a}_{2}+m\boldsymbol{a}_{1}$ and $\boldsymbol{a}^{(M)}_{2}=-m\boldsymbol{a}_{2}+(n+m)\boldsymbol{a}_{1}$, where both $m$ and $n$ being integers. The unit cell of the superlattice contains $N =4(n^2 + mn + m^2)$ atoms. The rotation angle $\theta$ is related to ($m,n$) by the following condition \cite{shall,shall2},
\begin{equation}
 \cos{(\theta)}=\frac{m^2+n^2+4mn}{2(m^2+n^2+mn)}.
 \label{eq:1}
\end{equation}
For $m=n=1$ ($m=1, n=0$), Eq.~(\ref{eq:1}) reduces to the well-known AA-stacked (AB-stacked) bilayer graphene. This implies that $\theta=0^\circ$ ($\theta=60^\circ$) corresponds to the perfect AA-stacking (AB-stacking).\par
Following the Ref.~\cite{tomanek}, we consider a Hamiltonian consisting of two parts: one is the intralayer part ($H_{\text{intra}}$), and the other is the interlayer part ($H_{\perp}$). This is true for two or more stacked arrangements of graphene layers. Hence, for a non-interacting system, the tight-binding Hamiltonian can be written in real space as (in the absence of any external field),
\begin{align}
H=&H_{\text{intra}}+H_{\perp} \nonumber \\ 
=& - \sum_{\substack{i\neq j \\ l}}t_{ij}^{ll}(c^{\dagger}_{l,i}c_{l,j} + \text{H.c.}) \nonumber
\\
&
-\sum_{\substack{i,j \\ l}}t_{ij}^{l,l+1}(c^{\dagger}_{l,i}c_{l+1,j} + \text{H.c.}).
\label{eq:2}
\end{align}
where the symbol $t_{ij}^{ll}$ in the first term $(H_{\text{intra}})$ represents the nearest neighbor intralayer hopping integral between sites $i$ and $j$. Specifically, $t_{ij}^{ll}$ takes on a value of $V_{pp\pi}^{0}=3.09$ eV, which acts at a distance $a_{0}=a/{\sqrt{3}}=0.142$ nm ($a_0$ is the carbon-carbon distance of monolayer graphene). $c^{\dagger}_{l,i}$ and $c_{l,i}$ denote the creation and the annihilation operators, respectively, at site $i$ within layer $l$. For a simple bilayer case, the layer index $l$ can take values of 1 or 2. H.c. denotes the hermitian conjugate term. Further, we have not considered any next-nearest neighbor intralayer hopping terms in $H_{\text{intra}}$.\par 
The second term, denoted as $H_{\perp}$ in Eq.~(\ref{eq:2}) describes the interaction between layers. For simplicity, we will first explain the term $H_{\perp}$ for an AB-stacked bilayer graphene. We consider two atoms placed on top of each other and separated by a distance $d_{0}$, then the interlayer hopping integral between these atoms becomes $t(0)=V_{pp\sigma}^{0}=0.39$ eV. The onsite energy is set to zero on all atoms. However, in the context of TBG, where the planes are rotated relative to each other, the interlayer hopping integral between sites having an in-plane projection of $r$ and an out-of-plane projection of $d_{0}$ can be expressed as follows,
\begin{equation}
t(r)=V_{pp\sigma}^{0}e^{-(\sqrt{r^2+d^2_0}-d_0)/\lambda}\frac{d^2_{0}}{r^2+d^2_{0}}
\label{eq:3}
\end{equation} 
Here, $V_{pp\sigma}$ remains the dominant factor, but it is adjusted by both the distance and the cosine of the twisted angle, as elaborated in Ref.~\cite{moon2}. $d_{0}$ is taken to be 3.35~\AA. $\lambda$ is the decay parameter that fine-tunes the cutoff of $t(r)$. In our calculations, we adopt a value of $\lambda=0.27$~\AA, which effectively reproduces similar band structures for bilayer graphene stacks with both AA and AB configurations.  
\section{formalism}\label{III}
In this section, we shall describe the numerical method used to compute the transport properties of a TBG system. Our approach is based on a real-space implementation of the Kubo formalism to calculate the dc conductivity of large systems, a technique developed by Garcia et al. \cite{rappoprt}. The numerical implementation of the Kernel Polynomial Method (KPM) \cite{weisse, simao3} entails the expansion of the Bastin formula in terms of Chebyshev polynomials for obtaining the conductivity tensors within the linear response regime. For non-interacting electrons, the components of the dc conductivity tensor ($\omega \rightarrow 0$) \cite{bastin,rappoprt,ortmann} can be expressed using the Kubo-Bastin formula \cite{kubo1, kubo2} as,
\begin{align}
\sigma_{\alpha\beta} (\mu, T) = & \frac{i e^2 \hbar}{A}\int_{-\infty}^{\infty} d\epsilon f(\epsilon) \text{Tr} \Big<v_{\alpha} \delta(\epsilon-H)v_\beta \frac{dG^{+}(\epsilon)}{d\epsilon} \nonumber
\\
&
-v_{\alpha} \frac{dG^{-}(\epsilon)}{d\epsilon}v_\beta\delta(\epsilon-H)\Big>,
\label{eq:4}
\end{align} 
where $\mu$ stands for the chemical potential, while $T$ denotes the temperature. The area of the sample is denoted by $A$. $v_\alpha$ and $v_\beta$ are the components of the velocity operator for $\alpha$ and $\beta$ respectively. $f(\epsilon)$ denotes the Fermi-Dirac distribution function for a given $\mu$ and $T$. $G^{\pm}(\epsilon, H)=1/(\epsilon-H\pm i\eta)$ represents the advanced ($`+$') and the retarded ($`-$') Green's function respectively.\par 
Using KPM, we can expand the spectral representation of both the rescaled delta and Green's function in terms of Chebyshev polynomials \cite{weisse}. Consequently, the conductivity tensor (Eq.~(\ref{eq:4})) can be written as, 
\begin{align}
\sigma_{\alpha\beta} (\mu, T) =& \frac{4 e^2 \hbar}{\pi A} \frac{4}{(\Delta E)^2} \int_{-1}^{1} d{\tilde{\epsilon}} \frac{f(\tilde{\epsilon})}{(1-{\tilde{\epsilon}^2})^2}
\sum_{m,n} \Gamma_{nm} (\tilde{\epsilon})
\mu_{nm}^{\alpha\beta} (\tilde{H})
\label{eq:5}
\end{align} 
where $\tilde{\epsilon}$ denotes the rescaled energy within the [$-$1, 1] range. Similarly, $\tilde{H}$ represents the rescaled Hamiltonian, while $\Delta E$ is the range of the energy spectrum. The functions $\Gamma_{nm} (\tilde{\epsilon})$ and $\mu_{nm}^{\alpha\beta} (\tilde{H})$ depend on the rescaled energy and Hamiltonian respectively, and can be expressed as, 
\begin{align}
 \Gamma_{nm} (\tilde{\epsilon})\equiv &(\tilde{\epsilon}-in\sqrt{1-\tilde{\epsilon}^2})e^{in~arccos(\tilde{\epsilon})} T_m(\tilde{\epsilon}) \nonumber
 \\
 &
+ (\tilde{\epsilon}+im\sqrt{1-\tilde{\epsilon}^2})e^{-im~arccos(\tilde{\epsilon})} T_n(\tilde{\epsilon}) 
\label{eq:6}
\end{align}
and 
\begin{equation}
\mu_{nm}^{\alpha\beta}(\tilde{H})\equiv \frac{g_m g_n}{(1+\delta_{n0})(1+\delta_{m0})} \text{Tr}\big[v_{\alpha}T_m(\tilde{H})v_{\beta}T_n(\tilde{H})\big]
\label{eq:7}
\end{equation}
where the latter involves the product of Chebyshev polynomial expansions. The Jackson kernel, denoted as $g_m$ is used to smoothen out the Gibbs oscillations \cite{weisse}, which arise due to truncation of the expansion outlined in Eq.~(\ref{eq:5}). $T_m(x)$ is the Chebyshev polynomials, defined by the following recurrence relation,
\begin{equation}
T_m(x)= 2x T_{m-1}(x)-T_{m-2}(x),
\label{eq:8}
\end{equation}
where $T_0(x)=1$ and $T_1(x)=x$.\par
The density of states (DOS) can be calculated using the spectral operator $\delta(\tilde{\epsilon}-\tilde{H})$ as,
\begin{equation}
\rho(\tilde{\epsilon})=\frac{1}{N} \text{Tr} \delta(\tilde{\epsilon}-\tilde{H})=\frac{1}{\pi\sqrt{1-{\tilde{\epsilon}}^2}}\sum_{n=0}^{\infty}\mu_{n}T_{n}(\tilde{\epsilon})
\label{eq:9}
\end{equation}
where $\mu_{n}$ is the Chebyshev moments and given by
\begin{equation}
\mu_{n}=\frac{1}{N} \frac{(1+\delta_{n,0})}{2}\text{Tr}~T_{n}(\tilde{H}).
\label{eq:10}
\end{equation}
\begin{figure*}[!ht!]
\begin{center}
		\phantomlabelabovecaption{(a)}{fig:2a}
		\phantomlabelabovecaption{(b)}{fig:2b}
		\phantomlabelabovecaption{(c)}{fig:2c}
\includegraphics[width=\textwidth]{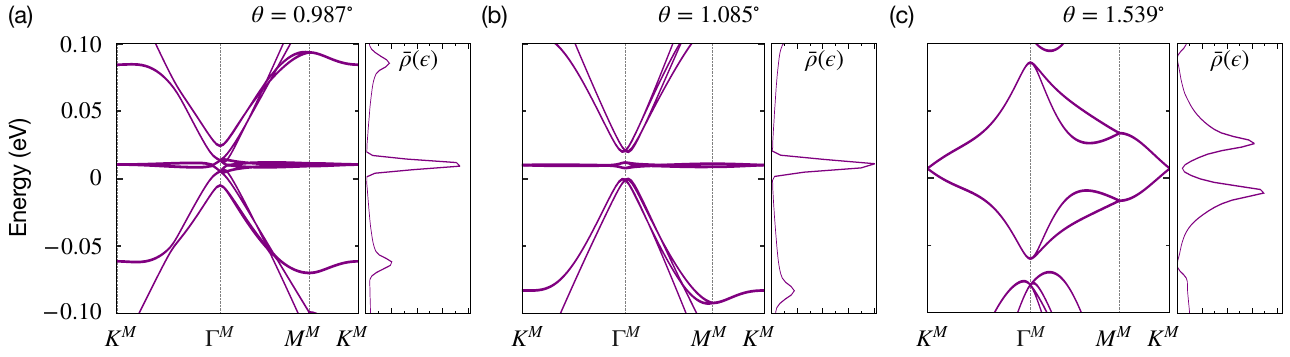}
\caption{(Color online) The electronic band dispersion along the high-symmetry points $K^{M}$ $\rightarrow$ $\Gamma^{M}$ $\rightarrow$ $M^{M}$ $\rightarrow$ $K^{M}$ and the corresponding normalized density of states, $\bar{\rho}(\epsilon)$ of TBG are shown for (a) $\theta= 0.987^\circ$ and ($m$, $n$) = (34, 33), (b) $\theta=1.085^\circ$ and ($m$, $n$) = (31, 30) and (c) $\theta=1.539^\circ$ and ($m$, $n$) = (22, 21) respectively.}
\label{fig:2}
\end{center}
\end{figure*}
\section{Results}\label{IV}
In the following sections, we will present our numerical results including the electronic and transport properties of the TBG system. We have simulated a very large system that consists of approximately atomic sites of the order of $10^7$ (ten million). Since the number of atoms in the unit cell increases with the decrease of the twist angle, we have adjusted the number of unit cells in both the $x$ and $y$ directions in order to maintain the total number of atoms within the system. Consequently, the computational costs are very much demanding near the magic angle. For our analysis, we have employed a truncation order of $M= 12000$ (Chebyshev moments) which is quite large and thus provides a reasonable level of accuracy. The convergence of the peaks depends on the number of $M$. Nevertheless, it's possible to manipulate the system size and increase the number of Chebyshev moments further to reduce fluctuations. We have also imposed periodic boundary conditions for all our numerical simulations.  
\subsection{Electronic structures (Flat bands)}\label{A}
In this section, we mainly focus on the electronic band structure, particularly emphasizing the flat band characteristics of a TBG system near the small twist angles at zero external field. The aim is to illustrate the behavioral transition observed below and above the magic angle in our system. Additionally, we present the density of states (DOS) plots for the same purpose. In Figs.~{\ref{fig:2a}-\ref{fig:2c}}, we calculate the energy band structures (left) and the corresponding normalized DOS, $\bar{\rho}(\epsilon)$ (right) within the low energy regime for three commensurate small twist angles, $\theta=0.987^\circ$, 1.085$^\circ$ and 1.539$^\circ$ respectively using Eq.~(\ref{eq:2}). Fig.~\ref{fig:2a} shows that the bands are almost flat in the vicinity of zero Fermi energy when $\theta=0.987^\circ$. This nearly flat band significantly contributes to the DOS for $\theta=0.987^\circ$ which can also be seen from Fig.~\ref{fig:2a}. The DOS exhibits a prominent sharp peak accompanied by two additional smaller van Hove peaks, which also correspond to the flatness of the band dispersion within the low energy regime. With the tuning of $\theta$, the energy bands exhibit enhanced flatness, leading to the formation of localized states with Fermi velocity close to zero value for $\theta=1.085^\circ$. This is depicted in Fig.~\ref{fig:2b}. At $\theta=1.085^\circ$, we observe our so-called ``magic angle" with two distinct gaps (one above and one below the Fermi energy). However, these gaps are now separated from the bulk bands with the reduced bandwidth as compared to $\theta=0.987^\circ$ where the band overlapping occurs (see Figs.~\ref{fig:2a} and \ref{fig:2b}). As a result, the DOS also shows a similar sharp peak near the zero Fermi energy when $\theta=1.085^\circ$, but the two other van Hove peaks associated with it are now shifted slightly towards the higher energy as compared to $\theta=0.987^\circ$ (see Figs.~\ref{fig:2a} and \ref{fig:2b}). These flat bands that arise due to the strong interlayer coupling also have implications in the LL spectra, which has been discussed later in Sec.~(\ref{C}).\par
Above the magic angle (when $\theta=1.539^\circ$), the flatness of the band near the zero Fermi energy almost vanishes and becomes dispersive within the same energy range, especially near the $\Gamma^{M}$ point in the Brillouin zone (see Fig.~\ref{fig:2c}). The corresponding DOS reveals the presence of two diverging van Hove singularities located near the Fermi level. However, the separation between these two singularities increases as the twist angle continues to increase \cite{brihuega}. It is important to note that the energy spectrum as well as the DOS is not symmetrical with respect to $E=0$ for small twist angles shown in Fig.~\ref{fig:2}. Hence we observe a small shift in Fermi energy due to broken particle-hole symmetry of the model at small twist angles. 
\begin{figure*}[!ht!]
\begin{center}
	\phantomlabelabovecaption{(a)}{fig:3a}
	\phantomlabelabovecaption{(b)}{fig:3b}
	\phantomlabelabovecaption{(c)}{fig:3c}
\includegraphics[width=\textwidth]{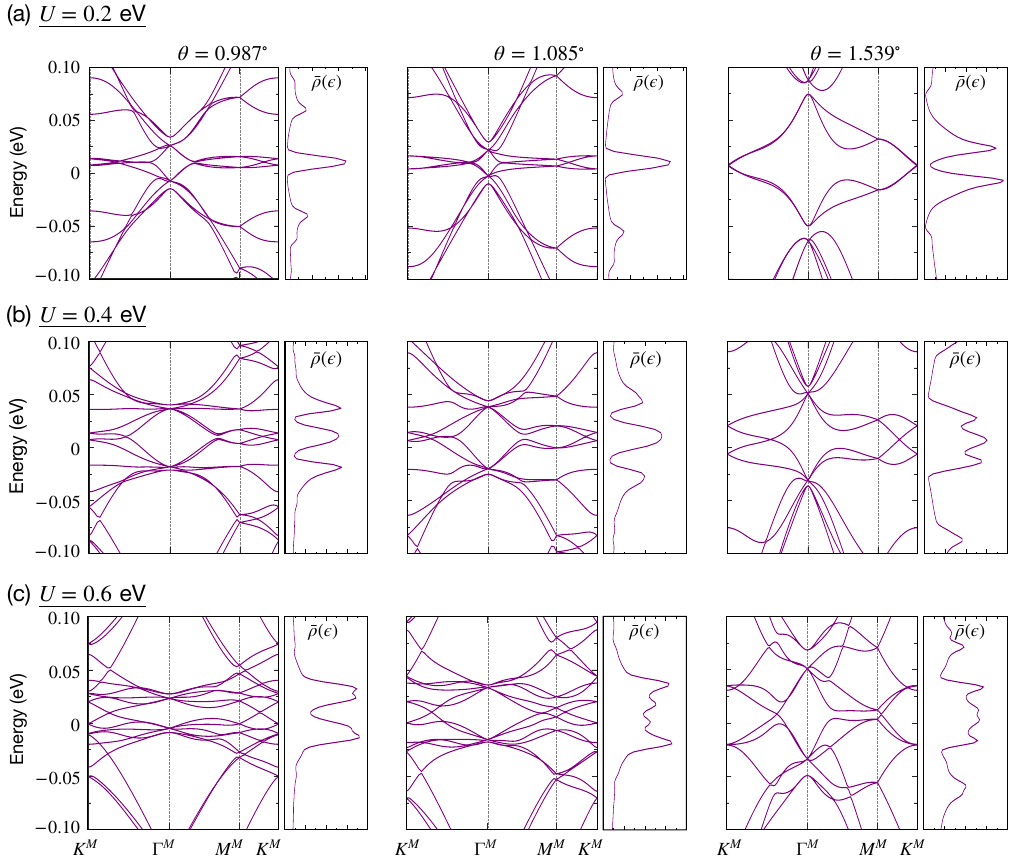}
\caption{(Color online) The electronic band dispersion along the high-symmetry points $K^{M}$ $\rightarrow$ $\Gamma^{M}$ $\rightarrow$ $M^{M}$ $\rightarrow$ $K^{M}$ and the corresponding normalized density of states, $\bar{\rho}(\epsilon)$ of TBG are shown for (a) $U=0.2$ eV, (b) $U=0.4$ eV and (c) $U=0.6$ eV with $\theta=0.987^\circ$ and ($m$, $n$) = (34, 33) (in the left panel), $\theta=1.085^\circ$ and ($m$, $n$) = (31, 30) (in the middle panel) and $\theta=1.539^\circ$ and ($m$, $n$) = (22, 21) (in the right panel) respectively.}
\label{fig:3}
\end{center}
\end{figure*}     
\subsection{Effects of an external Electric field}\label{B}
In this section, we discuss the effects of an external electric field (which breaks the inversion symmetry) on the electronic properties of the TBG system at $\theta=0.987^\circ$, 1.085$^\circ$ and 1.539$^\circ$. To introduce a perpendicular electric field, we apply a potential of $+U$/2 on layer 1 (lower) and $-U$/2 on layer 2 (upper). In Fig.~\ref{fig:3}, we show how the energy band structures and the normalized density of states, $\bar{\rho}(\epsilon)$ depend on the potential difference between the two layers with a small twist and also evolves with the tuning of the electric field strength. It is well-known that the Dirac cone touching is protected by the $C_{2}$T (time-reversal) symmetry and can be gapped out by breaking this symmetry. In the case of a TBG system, the external electric field does not induce a band gap in the spectrum protected by the $C_2$ (180$^\circ$) rotational symmetry \cite{moon}. Nevertheless, it does facilitate the transfer of electronic charge from the lower layer to the upper layer of the system \cite{morell}. This excess electronic charge has a strong variation for small angles.\par
In the absence of an external electric field, the two bands (namely, the lowest conduction band and the highest valence band) across the Fermi energy touch each other at the superlattice Dirac point labeled as $K^{M}$ (see Fig.~\ref{fig:2}). When an electric field $U$ is introduced, the degeneracy of the two layers (upper and lower) at the $K^{M}$ points is lifted owing to different onsite energies induced in each layer. As a consequence, the two Dirac points move in opposite directions in energy, leading to a gap opening exactly at the $K^{M}$ point of the moir\'e BZ. We denote it as $\Delta E(K^{M})$. We also find that the DOS peaks get broadened and eventually the singularity in DOS vanishes with the increasing field. Fig.~\ref{fig:3a} illustrates the band structures as well as the DOS at $U=0.2$ eV for $\theta=0.987^\circ$ (left panel), 1.085$^\circ$ (middle panel), and 1.539$^\circ$ (right panel) respectively. The flatness behavior of the energy bands near the zero Fermi energy observed for the $U=0$ case is now reduced with the lifting degeneracy, resulting in broadened peaks in the DOS. The broadening is a consequence of a dispersive splitting in the flat bands for $U \neq 0$. This phenomenon is evident in Figs.~\ref{fig:3a} for $\theta=0.987^\circ$ and $\theta=1.085^\circ$ respectively. Additionally, the gap above and below the Fermi energy is no longer observed for $\theta=1.085^\circ$ in the presence of the electric field. However, this effect is negligible for $U=0.2$ eV when $\theta=1.539^\circ$, as shown in the right panel of Fig.~\ref{fig:3a}. When the strength of the electric field, $U$ is increased from 0.2 to 0.4 eV, the band dispersion and the corresponding DOS starts evolving in a peculiar fashion below a certain $\theta$ value. The energy difference within the van Hove peaks in the DOS around the zero Fermi energy diminishes as the flat bands disappear, as depicted in Fig.~\ref{fig:3b}. As we further increase the electric field value (say, $U=0.6$ eV), we observe the significant contribution from the bulk bands especially near the Fermi energy where the bands are dispersive (see Fig.~\ref{fig:3c}). This underscores the possibility of controlling the position of both the Fermi energy and the VHS by varying the twist angle and gate voltage, respectively. This offers a potent toolkit for manipulating the electronic states of the system. \par 
For clarity of our findings, we further investigate the tunability of this $K^{M}$ point gap, denoted as $\Delta{E}(K^M)$ as a function of both rotation angle, $\theta$ and the electric field strength, $U$. In Fig.~\ref{fig:4a}, we have shown the energy difference, $\Delta{E}(K^M)$ while varying the twist angle, $\theta$ for $U=0.2$, 0.4 and 0.6 eV. Interestingly, $\Delta{E}(K^M)$ remains almost constant for a certain range of $\theta$ ($\theta>10^{\circ}$) across all the values of $U$, converging to $U$ as $\Delta{E}(K^M) \rightarrow U$, and subsequently exhibits a steep decline for $\theta<10^{\circ}$. Nonetheless, this feature becomes particularly remarkable at extremely small twist angles depending on the values of the external electric field, as depicted in Fig.~\ref{fig:4b} with a zoomed view. For the previous case ($\theta>10^{\circ}$), the shift occurs at +$U$/2 for layer 1 in the positive energy and at $-U$/2 for layer 2 in the negative energy owing to the weak interlayer coupling near zero Fermi energy where the layers behave as decoupled. However, for the latter case, we observe multiple closing and re-opening of the $K^{M}$ point gap approximately for $\theta<3^{\circ}$ since the energy levels disperse non-monotonically with the variation of the electric field strength. At low field strength ($U=0.2$ eV), this anomalous behavior of the energy gap manifests below $1.6^{\circ}$, as indicated by the black curve whereas for the other two values, $U=0.4$ and 0.6 eV, the same phenomenon occurs below $1.8^{\circ}$ (indicated by the blue curve) and $2.4^{\circ}$ (indicated by the brown curve), respectively. Fig.~\ref{fig:4c} shows the plot for $\Delta{E}(K^M)$ as a function of $U$ when $\theta=1.085^{\circ}$. The non-monotonic behavior in $\Delta{E}(K^M)$ is still observed with a similar gap closing and re-opening while varying the field strength $U$.  
\begin{figure}[t]
\begin{center}
	\phantomlabelabovecaption{(a)}{fig:4a}
	\phantomlabelabovecaption{(b)}{fig:4b}
	\phantomlabelabovecaption{(c)}{fig:4c}
\includegraphics[width=0.4\textwidth,height = 16cm]{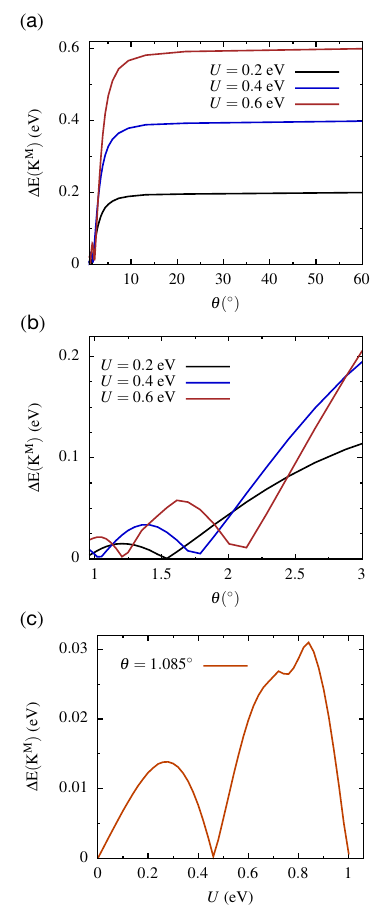}
\caption{(Color online) (a) Energy gap, $\Delta{E}(K^M)$ (in units of eV) is shown as a function of twist angles, $\theta$ (in units of degree) for $U=0.2$, 0.4, and 0.6 eV (b) Zoomed view of $\Delta{E}(K^M)$ near the small twist angle ($1^\circ \leq \theta \leq 3^\circ$) is shown. (c) Energy gap, $\Delta{E}(K^M)$ (in units of eV) is shown as a function of electric field strength, $U$ (in units of eV) at $\theta=1.085^{\circ}$.}
\label{fig:4}
\end{center}
\end{figure}
\subsection{Effects of the magnetic field}\label{C}
In this section, we shall explore the effects of a magnetic field applied perpendicular to the plane of a TBG system in the absence of an external electric field. We consider a uniform magnetic field $\vec{B}=B\hat{z}$, where $B$ represents the strength of the magnetic field. We incorporate the magnetic field effect by multiplying the Peierls phase factor $e^{2 i\pi\phi_{ij}}$ (where $\phi_{ij}$ is the magnetic flux) into the hopping terms as outlined in Eq.~(\ref{eq:2}) using the Peierls substitution method \cite{peierls,ezawa,sinha}. It is important to note that, in contrast to a conventional superlattice, the strength of the magnetic field, $B$ in a TBG system is determined by the size of the unit cell. As the dimensions of the unit cell increase, the necessary strength of the magnetic field decreases. Consequently, with smaller twist angles which correspond to larger unit cell dimensions, a weaker magnetic field is sufficient to quantize the single-particle electronic states into LL \cite{do}. \par
\begin{figure}[!t!]
\begin{center}
	\phantomlabelabovecaption{(a)}{fig:5a}
	\phantomlabelabovecaption{(b)}{fig:5b}
\includegraphics[width=0.42\textwidth]{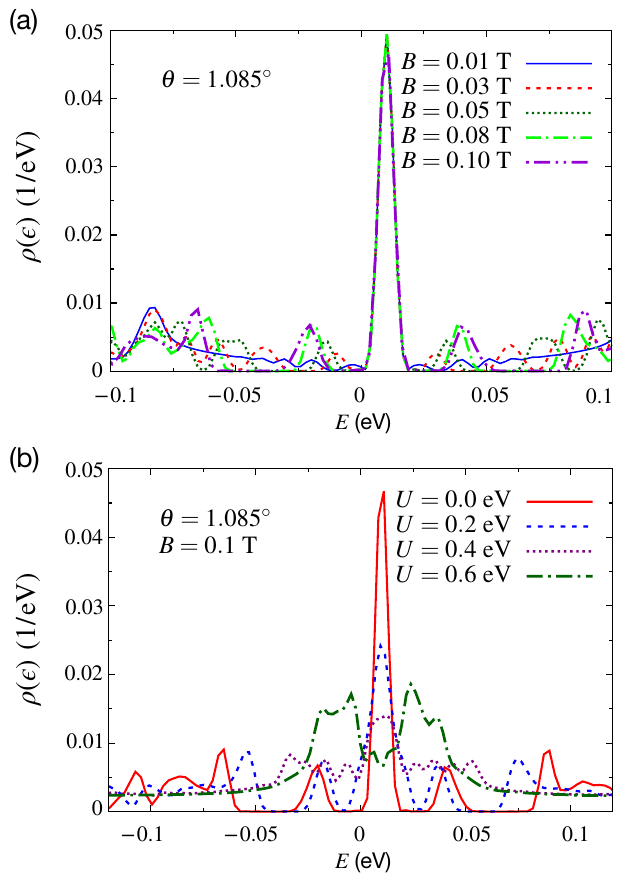}
\caption{(Color online) (a) The density of states, $\rho(\epsilon)$ (in units of 1/eV) is shown as a function of Fermi energy, $E$ (in units of eV) for different values of $B$ at $\theta=1.085^\circ$. (b) The density of states, $\rho(\epsilon)$ (in units of 1/eV) is plotted as a function of Fermi energy, $E$ (in units of eV) for different values of electric field strength, $U=0.0$, 0.2, 0.4, and 0.6 eV at a fixed value of $B$ ($B=0.1$ T) when $\theta=1.085^\circ$.}
\label{fig:5}
\end{center}
\end{figure}
\begin{figure}[!ht!]
\begin{center}
	\phantomlabelabovecaption{(a)}{fig:6a}
	\phantomlabelabovecaption{(b)}{fig:6b}
\includegraphics[width=0.45\textwidth]{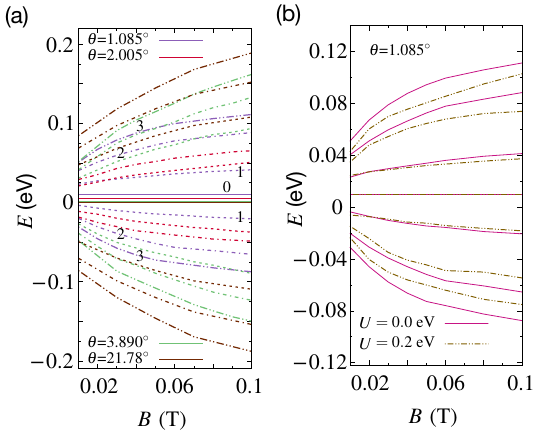}
\caption{(Color online) (a) Landau level energies, $E$ (in units of eV) are plotted as a function of the magnetic field strength, $B$ (in units of T) for different twist angles, $\theta=1.085^\circ$, 2.005$^\circ$, 3.890$^\circ$, 21.78$^\circ$ as indicated by the different colors. The solid lines indicate the zero-mode energy states for all the twist angles. $n=1, 2, 3$ denotes the three lowest Landau levels with index $n$ for $\theta$=1.085$^\circ$. (b) Landau level energies, $E$ (in units of eV) are shown as a function of the magnetic field strength, $B$ (in units of T) for $U=0$ and 0.2 eV at $\theta=1.085^\circ$.}
\label{fig:6}
\end{center}
\end{figure}
In Fig.~\ref{fig:5a}, we numerically calculate the density of states, $\rho(\epsilon)$ as a function of Fermi energy, $E$ for different values of $B$ with $\theta$=1.085$^\circ$. At a very low magnetic field ($B=0.01$ T), the discreteness of the LL peaks is very small. However, as we increase the magnetic field value $B$, the LL peaks become more pronounced. Nevertheless, only a few LL can be clearly distinguished away from the charge neutrality point, even under a stronger magnetic field ($B=0.1$ T) for $\theta=1.085^\circ$ which is not true for large twist angles. The influence of the higher LL is overridden by those originating from the upper and lower boundaries of the energy bands. Most importantly, the DOS near the zero Fermi energy, corresponding to the flat band, is independent of the strength of the magnetic field since it has topological protection \cite{gail}. Furthermore, we illustrate the LL spectra, labeled as $E$, as a function of the magnetic field $B$  for a range of twist angles, specifically 1.085$^\circ$, 2.005$^\circ$, 3.89$^\circ$ and 21.78$^\circ$ within the extremely low magnetic field regime and at low energies. These results are depicted in Fig.~\ref{fig:6a}. When the twist angle is sufficiently large (as in the case of $\theta=21.78^\circ$), the low-energy LL spectrum behaves similarly to monolayer graphene, with a dependence on both $n$ ($n>0$) and $B$ following a square-root relationship. This suggests that, at lower energies, interlayer interactions are relatively weak. However, as we reduce the twist angle, for example, to $\theta=3.89^\circ$ and 2.005$^\circ$,  we still observe similar structures resembling the Landau levels of monolayer graphene in the low-energy LL spectrum (below 0.2 eV), albeit with variations in the energy scale. When the twist angle becomes even smaller, as in the case of a magic angle ($\theta=1.085^\circ$), the low energy LL spectrum becomes compressed near the Dirac points. This behavior is consistent with the observed reduction in bandwidth for smaller twist angles due to significant interlayer coupling, in contrast to larger twist angles. However, the zero energy level remains flat and nondispersive which is true for all the twist angles (see Fig.~\ref{fig:6a}). Additionally, the zero energy peaks shift towards positive energies (electron side) as $\theta$ decreases. There are energy level crossings as a function of the magnetic field for various twist angles within this low field regime. However, the difference between the zeroth mode and the immediately higher or lower LL narrows as twist angles reduce. When the LL energy falls below the van Hove energy (0.1 eV), energy levels exhibit eight-fold degeneracy (spin, valley, and layers degree of freedom) at very low magnetic fields, reducing to four-fold degeneracy at higher fields.
\begin{figure}[!ht!]
\begin{center}
\subfloat{\includegraphics[width=0.45\textwidth]{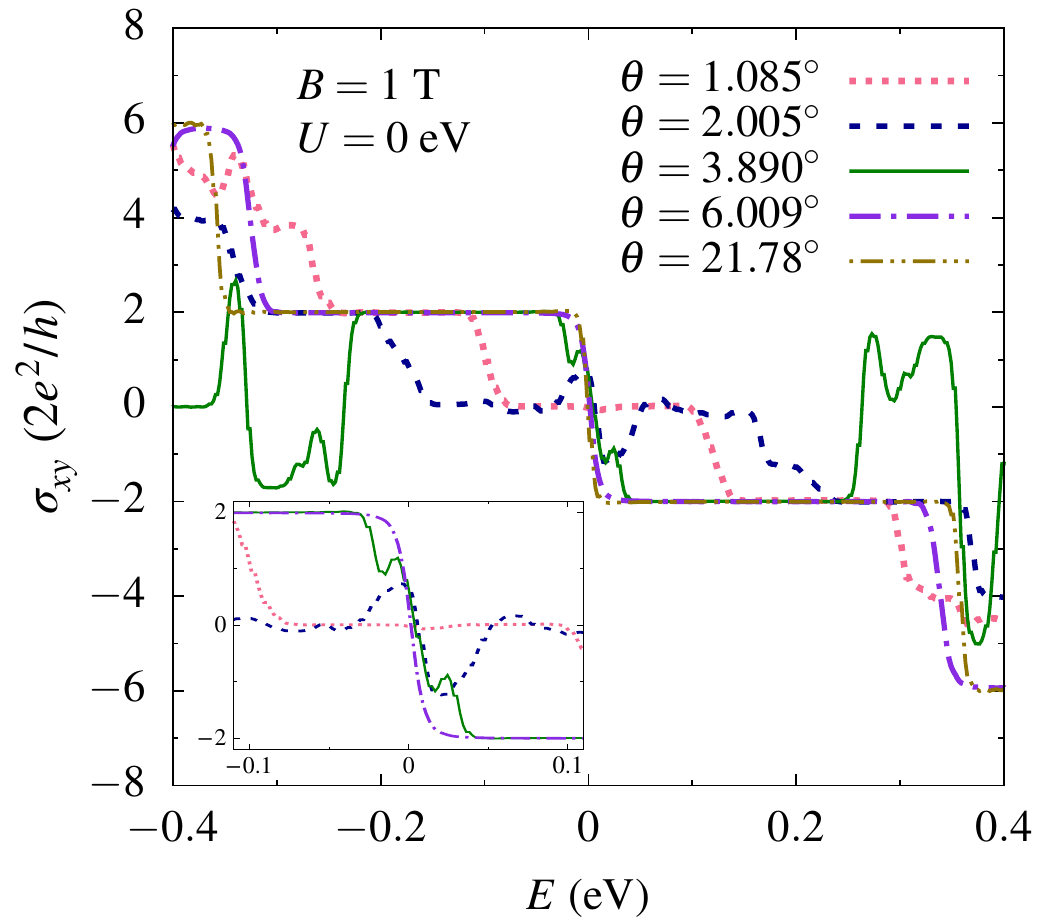}}
\caption{(Color online) Hall conductivity, $\sigma_{xy}$ (in units of $2e^2/h$) is shown as a function of Fermi energy, $E$ (in units of eV) for different twist angles, namely $\theta=1.085^\circ$, 2.005$^\circ$, 3.89$^\circ$, 6.009$^\circ$, and 21.78$^\circ$ for $B=1$ T and $U=0$ eV.}
\label{fig:7}
\end{center}
\end{figure}
\subsection{Effects of both electric field and magnetic field}\label{D}
In this section, we explore the influence of an external perpendicular electric field, $U$ on the LL spectra in a TBG system at low magnetic fields. We first observe how the LL peaks get modified while varying the strength of the electric field. Fig.~\ref{fig:5b} shows the DOS, $\rho(\epsilon)$ as a function of Fermi energy, $E$ at low $B$ field ($B=0.1$ T) for $U= 0$, 0.2, 0.4 and 0.6 eV with $\theta=1.085^\circ$. Within the low energy range, we observe a reduction in the number of LL peaks as the electric field strength increases. Eventually, at a critical electric field strength ($\approx$ 0.6 eV), the LL peaks collapse as shown in Fig.~\ref{fig:5b}. Consequently, it becomes possible to achieve electrically adjustable LL spectra in TBG. Fig.~\ref{fig:6b} shows the LL spectra as a function of the magnetic field, $B$ under an electric field strength of $U=0.2$ eV. For the sake of comparison, we have also included the LL corresponding to the case where $U=0$ eV. Notably, the zero energy peak remains unchanged for both scenarios, i.e., for $U=0$ eV and $U\neq0$ eV. However, the higher energy LL ($n>0$) exhibit a discernible shift towards lower energies for the $U\neq0$ case, as indicated by the dotted lines in Fig.~\ref{fig:6b}. Moreover, we observe the energy level crossings in the LL spectra at very low energies. 
\begin{figure*}[!ht!]
\begin{center}
	\phantomlabelabovecaption{(a)}{fig:8a}
	\phantomlabelabovecaption{(b)}{fig:8b}
\includegraphics[width = \textwidth]{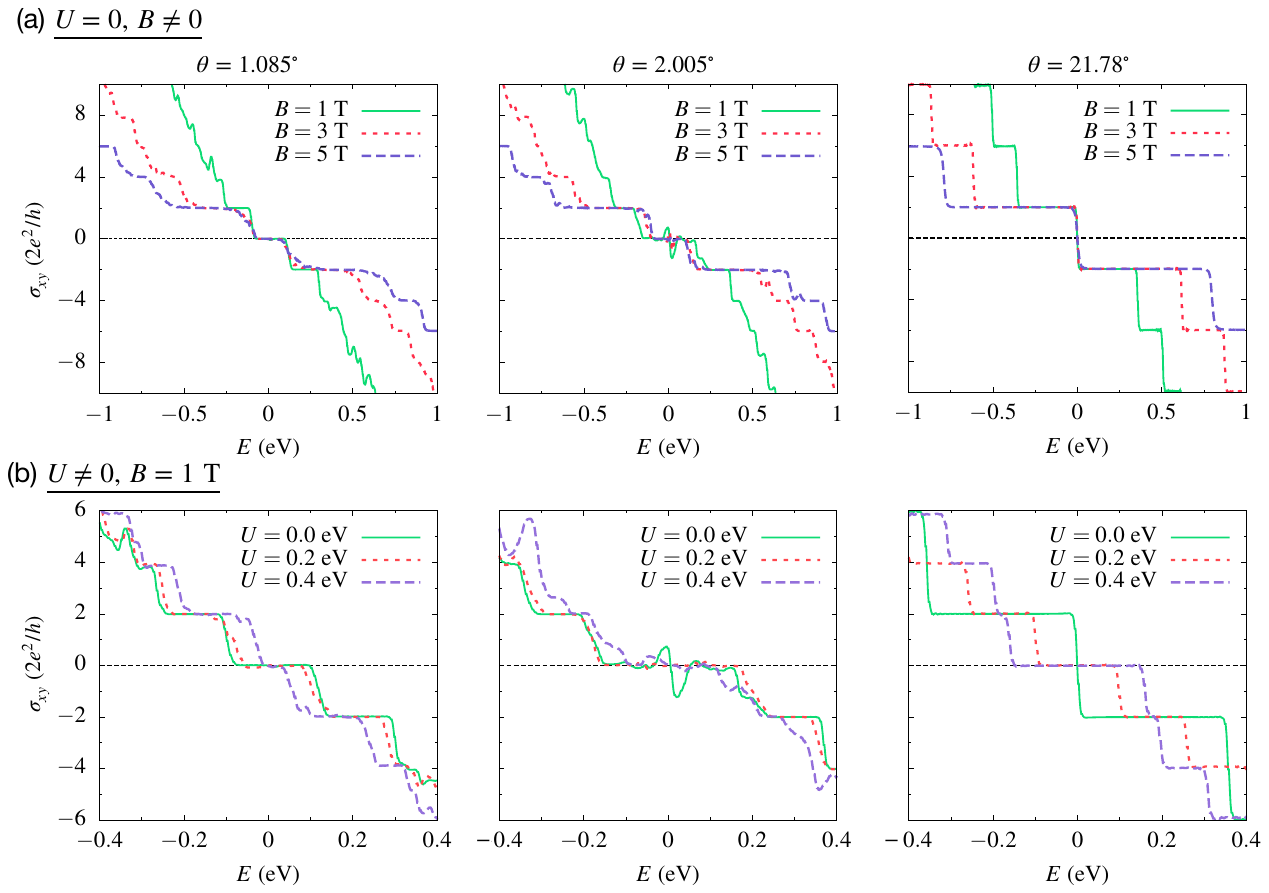}
\caption{(Color online) Hall conductivity, $\sigma_{xy}$ (in units of $2e^2/h$) is plotted as a function of Fermi energy, $E$ (in units of eV) with (a) $U=0$, $B\neq 0$ in the top panel and (b) $U \neq0$, $B=1$ T in the bottom panel for $\theta=1.085^\circ$, $\theta=2.005^\circ$, and $\theta=21.78^\circ$ respectively.}
\label{fig:8}
\end{center}
\end{figure*}
\subsection{Hall conductivity}\label{E}
In this section, we investigate the transport properties of the TBG system in terms of the Hall conductivity, denoted as $\sigma_{xy}$, with respect to the Fermi energy, $E$ for different twist angles using Eq.~(\ref{eq:4}) as elaborated in Sec.~(\ref{III}). In Fig.~\ref{fig:7}, we have plotted $\sigma_{xy}$ as a function of Fermi energy, $E$ for $\theta=1.085^\circ$, 2.005$^\circ$, 3.89$^\circ$, 6.009$^\circ$, and 21.78$^\circ$ with $U=0$ eV and $B=1$ T at absolute zero temperature. The Hall conductivity, $\sigma_{xy}$ exhibits a sequence of quantized plateaus for all the twist angles but with a significant difference in the quantized value of $\sigma_{xy}$. More precisely, as we tune the twist angle from larger value to a smaller one, a critical point emerges around $\theta=2.005^\circ$ where $\sigma_{xy}$ shifts from a half-integer to an integer quantum Hall effect, \textit{i.e} $\sigma_{xy}= \pm 4(n+1/2)~(2e^2/h$) shifts to $\sigma_{xy}= \pm 2n~(2e^2/h$), n being the integer numbers. This is depicted in the inset of Fig.~\ref{fig:7}. Hence, the Hall quantization rules follow a different sequence with the variation of twist angles. \par 
Further, our findings include the results obtained for two distinct scenarios, (a) $U=0$, $B\neq0$ and (b) $U \neq 0$, $B\neq0$ as illustrated in Fig.~\ref{fig:8}. For $U=0$ and $B \neq0$, we have plotted $\sigma_{xy}$ as a function of Fermi energy, $E$ for different $B$ values ($B=1$ T, 3 T, and 5 T) with $\theta=1.085^\circ$, 2.005$^\circ$, 21.78$^\circ$ respectively as shown in Fig.~\ref{fig:8a} (top panel). For a very large twist angle (when $\theta=21.78^\circ$), $\sigma_{xy}$ follows a well-established sequence $\pm$2, $\pm$6, $\pm$10, $\ldots$ in units of $2e^2/h$ (2 is accounted for spin degeneracy). This Hall quantization is exactly double of a monolayer graphene having a linear band dispersion near the Dirac points. The underlying reason is obvious that the two layers behave as a decoupled layer, implying that the Landau level spectra are almost equivalent to the monolayer’s LL. As we reduce the twist angle to a lower value ($\theta=2.005^\circ$), the system undergoes a transition where a zero energy plateau emerges near the Dirac point with small fluctuations which arise due to the increased number of LL. Moreover, the Hall plateaus follow a different sequence as 0, $\pm$2, $\pm$4, $\pm$6, $\ldots$ in units of $2e^2/h$ when compared to the case at $\theta=21.78^\circ$. However, at the magic angle ($\theta=1.085^\circ$), the Hall plateaus follow the same sequence as observed for $\theta=2.005^\circ$ except for the fluctuations near the Dirac points. Furthermore, the width of these plateaus varies with increasing magnetic field strength $B$ for every twist angle. Additionally, the Hall plateaus exhibit some fluctuations away from the zero Fermi energy at higher energies for $\theta=1.085^\circ$ and $2.005^\circ$, which is absent for $\theta=21.78^\circ$. However, these artifacts can be improved with a higher truncation order of the expansion and by considering a larger sample size for the former case.\par
For the other scenario ($U \neq 0$, $B\neq0$), we show similar plots by varying the electric field strengths, $U=0$, 0.2, and 0.4 eV with $B=1$ T in Fig.~\ref{fig:8b}. In the presence of an electric field, we observe new quantized Hall plateaus in addition to the existing ones for $\theta=21.78^\circ$. Hence, the Hall quantization follows a different sequence as 0, $\pm$2, $\pm$4, $\pm$6, $\ldots$ in units of $2e^2/h$ as compared to the case when $U=0$. These steps of $4e^2/h$ between each plateau in $\sigma_{xy}$ indicate the presence of fourfold degeneracy of the LL for $U \neq 0$. This Hall conductivity plateau which exists at zero Fermi energy also supports the absence of zero energy LL (see Fig.~\ref{fig:10c} in Appendix). Moreover, the width of the plateaus changes asymmetrically as a function of Fermi energy when $U\neq0$. As we decrease the twist angle to $\theta=2.005^\circ$, the Hall conductivity shows a similar quantization rule (0, $\pm$2, $\pm$4, $\pm$6, $\ldots$ in units of $2e^2/h$) except the change in Hall plateau at zero Fermi energy as shown in Fig.~\ref{fig:8b}. Even, at the magic angle ($\theta=1.085^\circ$), the Hall quantization appears at the same value as observed previously for the $U=0$ case but with the decreasing width of the Hall plateau near the Dirac points. 
\section{Conclusions}\label{V}
In summary, we have studied the electronic and transport properties under the influence of external electric and magnetic fields employing a tight-binding model for a TBG system. In the absence of an external field, the energy bands of TBG near the Fermi level exhibit a flat nature,  leading to the localization of electronic states for small twist angles falling within the range of 0.985$^\circ<\theta<1.539^\circ$. However, when we apply the electric field, the degeneracy at the superlattice Dirac points gets lifted, and the energy gap behaves non-monotonically with the variation of twist angles, especially when $\theta<3^\circ$. This suggests that the electric field can be used as an effective tool to manipulate the electronic band structure in the proximity of the flat band region. Next, we explore how the LL spectra behave for different twist angles in a weak magnetic field. Moreover, we have demonstrated the behaviors of the LL spectra in the presence of an electric field. Our study of Landau levels confirms that LL peaks collapse at a certain critical electric field. This work opens up an opportunity for deciphering the interplay between external fields and the twisting effect on TBG. Additionally, we have found that the Hall quantization rules depend on the twist angle. At small angles, the Hall plateaus follow a different sequence as 0, $\pm$2, $\pm$4, $\pm$6, $\ldots$ in units of $2e^2/h$ as compared to the larger twist angles. Interestingly, at small twist angles, the electric field effect does not change the Hall quantization which is highly in contrast to the large twist angle. This study reveals the rich physics of TBG in the presence of uniform electric and magnetic fields. Moreover, it will contribute to the experimental investigations of magic angles with flat bands.
\begin{figure*}[!ht!]
\begin{center}
\subfloat[]{\includegraphics[width=0.33\textwidth]{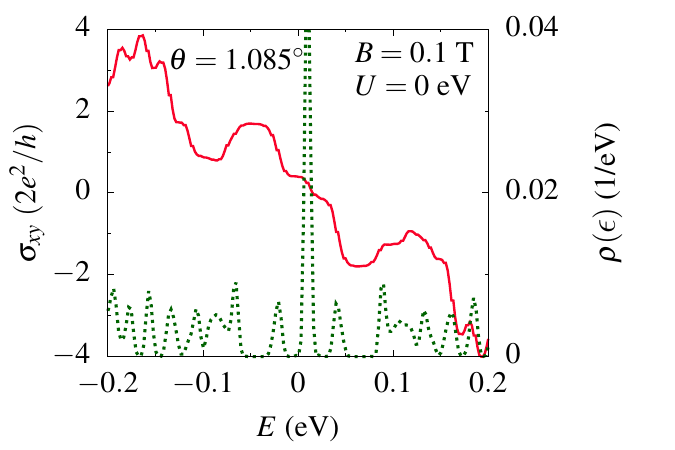}\label{fig:9a}} 
\subfloat[]{\includegraphics[width=0.33\textwidth]{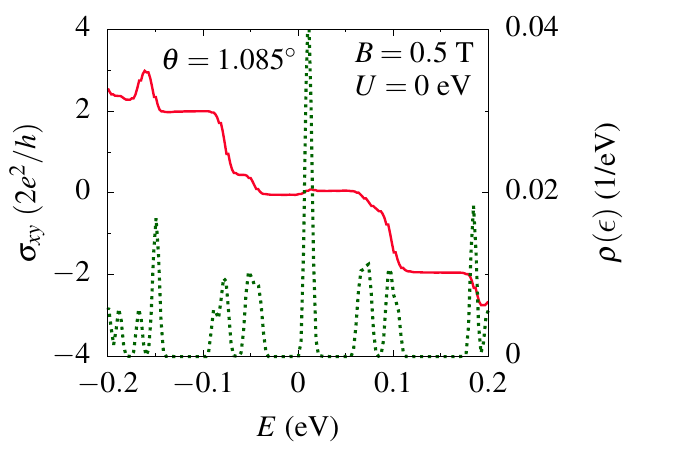}\label{fig:9b}}
\subfloat[]{\includegraphics[width=0.33\textwidth]{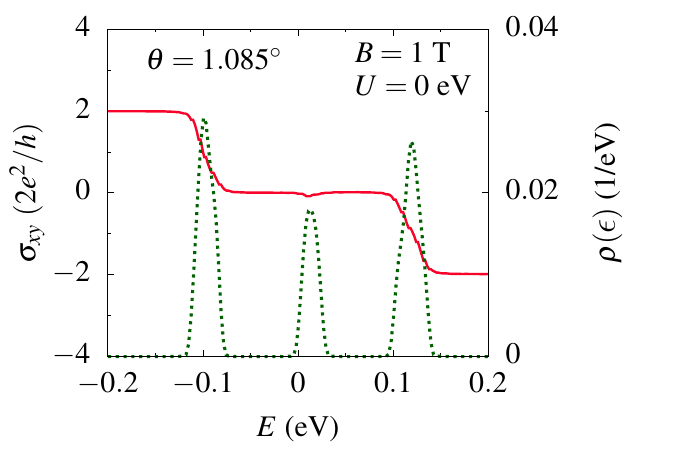}\label{fig:9c}}\\
\subfloat[]{\includegraphics[width=0.45\textwidth]{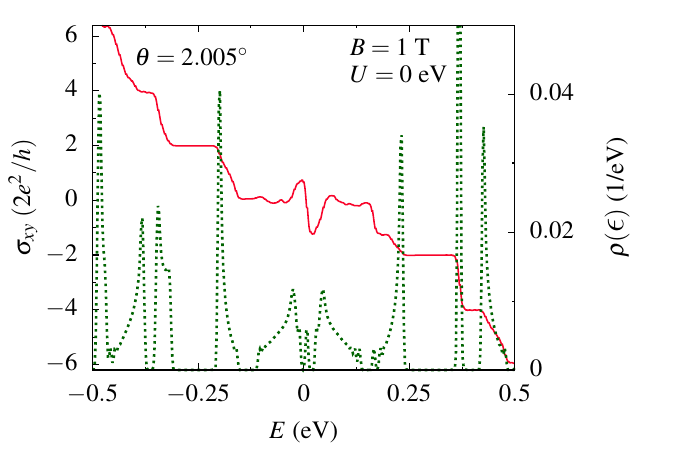}\label{fig:9d}}
\subfloat[]{\includegraphics[width=0.45\textwidth]{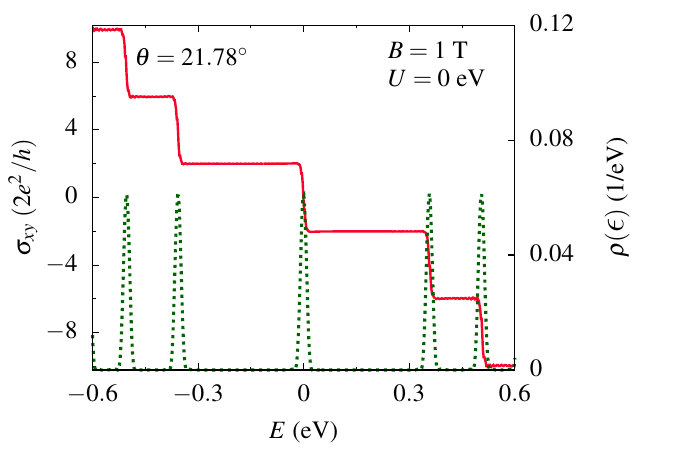}\label{fig:9e}}
\caption{(Color online) Hall conductivity, $\sigma_{xy}$ (in units of $2e^2/h$) and the density of states, $\rho(\epsilon)$ (in units of 1/eV) are plotted as a function of Fermi energy, $E$ (in units of eV) for (a) $B=0.1$ T, (b) $B=0.5$ T, and (c) $B=1$ T with $U=0$ eV at $\theta=1.085^\circ$. The same plots are shown for (d) $\theta=2.005^\circ$, and (e) $\theta=21.78^\circ$ with $B=1$ T and $U=0$ eV.}
\label{fig:9}
\end{center}
\end{figure*}
\begin{figure*}[!ht!]
\begin{center}
\subfloat[]{\includegraphics[width=0.33\textwidth]{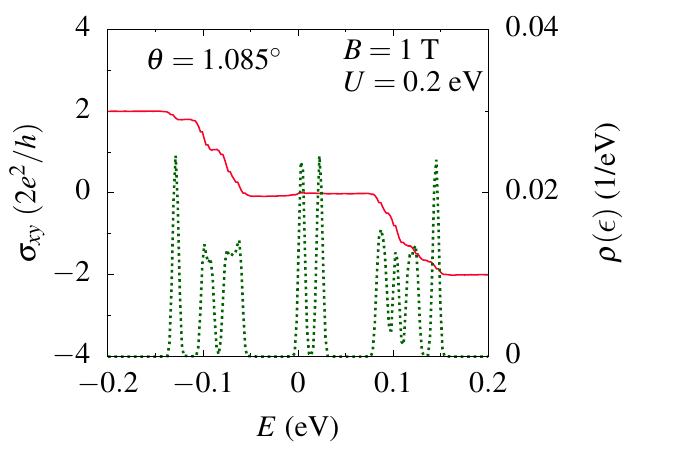}\label{fig:10a}} 
\subfloat[]{\includegraphics[width=0.33\textwidth]{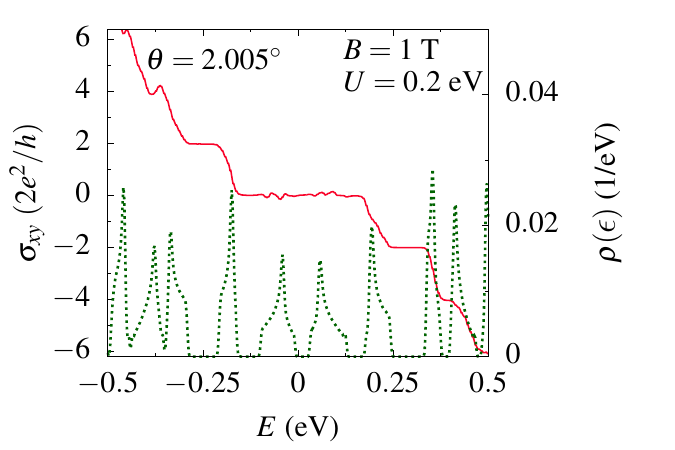}\label{fig:10b}}
\subfloat[]{\includegraphics[width=0.33\textwidth]{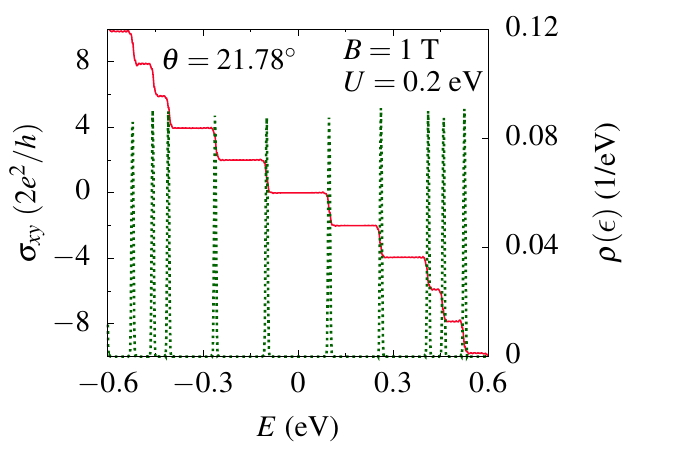}\label{fig:10c}}
\caption{(Color online) Hall conductivity, $\sigma_{xy}$ (in units of $2e^2/h$) and the density of states, $\rho(\epsilon)$ (in units of 1/eV) are plotted as a function of Fermi energy, $E$ (in units of eV) for (a) $\theta=1.085^\circ$, (b) $\theta=2.005^\circ$, and (c) $\theta=21.78^\circ$ with $B=1$ T and $U=0.2$ eV.}
\label{fig:10}
\end{center}
\end{figure*}
\begin{acknowledgments}
The authors acknowledge the support provided by the KEPLER computing facility, maintained by the Department of Physical Sciences, IISER Kolkata. P.~S.~and A.~M.~acknowledge the financial support from IISER Kolkata through the institute Post-doctoral and PhD Fellowship respectively. S.~M.~J.~acknowledges funding from the Royal Society through a Royal Society University Research Fellowship URF\textbackslash{}R\textbackslash{}191004. B.~L.~C.~acknowledges the SERB for grant no.~SRG/2022/001102 and ``IISER Kolkata Start-up-Grant'' Ref.No.IISER-K/DoRD/SUG/BC/2021-22/376. 
\end{acknowledgments}
\section*{Appendix: Landau levels versus Hall conductivity}
We demonstrate the behavior of the LL peaks, denoted as $\rho(\epsilon)$ during the transition of Hall conductivity, $\sigma_{xy}$ from one plateau to another with $\theta=1.085^\circ$, $2.005^\circ$, and $21.78^\circ$. We particularly focus on two different cases: (a) $U=0$, and (b) $U \neq 0$ as illustrated in Figs.~\ref{fig:9} and \ref{fig:10} respectively. In Figs.~\ref{fig:9a}-\ref{fig:9c}, we have shown the evolution of $\sigma_{xy}$ and $\rho(\epsilon)$ as a function of Fermi energy, $E$ for different magnetic field values (say, $B=0.1$, 0.5, and 1 T) with $\theta=1.085^\circ$ within the low energy range. Interestingly, we observe that there exists a zero energy peak in the DOS for all values of $B$ arising from the flat band characteristic. However, this zero energy peak does not contribute to the Hall conductivity, and consequently, a Hall plateau is present near the Dirac point. At a very low magnetic field (e.g., $B=0.1$ T), the Hall quantization values exhibit a non-discrete behavior as can be seen from Fig.~\ref{fig:9a}. Nonetheless, the overall shape of the plateau transitions corroborates with the characteristics of the LL peaks, as demonstrated in Fig.~\ref{fig:9a}. Upon increasing the strength of the magnetic field to $B=0.5$ T, the discreteness of the Hall plateaus becomes evident, and gets even more pronounced when $B=1$ T as shown in Figs.~\ref{fig:9b} and \ref{fig:9c} respectively. However, for $\theta=2.005^\circ$, we observe an increased number of LL peaks near the zero Fermi energy which is supported by the non-discrete behavior of the quantized Hall plateau near the zero Fermi energy as shown in Fig.~\ref{fig:9d}. However, at higher energies, the width of the plateaus changes with the difference between the two LL energies. For the large twist angle (when $\theta=21.78^\circ$), the gap between the two nearest neighbor LL decreases at higher energies as compared to the vicinity of the Dirac point. Hence the width of the plateau associated with the LL transitions also gets smaller as shown in Fig.~\ref{fig:9e}. Additionally, the presence of a zero energy LL peak indicates the absence of a zero energy Hall plateau.\par
In Figs.~\ref{fig:10a}-\ref{fig:10c}, we show the similar plots for $\theta=1.085^\circ$, $2.005^\circ$, and $21.78^\circ$ at $B=1$ T and $U=0.2$ eV. For $U \neq 0$, the LL peaks show splitting at $\theta=1.085^\circ$ while $\sigma_{xy}$ makes a transition from one plateau to another within the low energy range as shown in Fig.~\ref{fig:10a}. Moreover, the splitting is also observed in the DOS even when $\sigma_{xy}=0$ due to the broken layer degeneracy. As we increase the twist angle $\theta$ to $2.005^\circ$, the LL peaks splitting is still there in the DOS with $\sigma_{xy}=0$ as shown in Fig.~\ref{fig:10b}. However, for $\theta=21.78^\circ$, we observe the emergence of new plateaus in the Hall conductivity phenomenon, resulting in an increased number of LL peaks within the same energy range. This behavior is illustrated in Fig.~\ref{fig:10c}.   


\begin{thebibliography}{99}
\bibitem{wallace} P. R. Wallace, \href{https://doi.org/10.1103/PhysRev.71.622}{Phys. Rev. {\bf 71}, 622 (1947)}.

\bibitem{neto} A. H. Castro. Neto, F. Guinea, N. M. R. Peres, K. S. Novoselov, and A. K. Geim, \href{https://doi.org/10.1103/RevModPhys.81.109}{Rev. Mod. Phys. {\bf 81}, 109 (2009)}.

\bibitem{geim} A. K. Geim and K. S. Novoselov, \href{https://doi.org/10.1038/nmat1849}{Nat. Mater. {\bf 6}, 183 (2007)}.

\bibitem{geim2} A. K. Geim and I. V. Grigorieva, \href{https://doi.org/10.1038/nature12385}{Nature {\bf 499}, 419 (2013)}.

\bibitem{santos} J. M. B. Lopes dos Santos, N. M. R. Peres, and A. H. Castro Neto, \href{https://doi.org/10.1103/PhysRevLett.99.256802}{Phys. Rev. Lett. {\bf 99}, 256802 (2007)}.

\bibitem{Bistritzer} R. Bistritzer and A. H. MacDonald, \href{https://doi.org/10.1073/pnas.1108174108}{Proc. Natl. Acad. Sci. U.S.A. {\bf 108}, 12233 (2011)}.

\bibitem{chakov} A. O. Sboychakov, A. L. Rakhmanov, A. V. Rozhkov, and Franco Nori, \href{https://doi.org/10.1103/PhysRevB.92.075402}{Phys. Rev. B {\bf 92}, 075402 (2015)}.

\bibitem{Andrei} E. Y. Andrei and A. H. MacDonald, \href{https://doi.org/10.1038/s41563-020-00840-0}{Nat. Mater. {\bf 19}, 1265 (2020)}.

\bibitem{kim} K. Kim, A. Walter, L. Moreschini \textit{et al.}, \href{https://doi.org/10.1038/nmat3717}{Nat. Mater. {\bf 12}, 887 (2013)}.

\bibitem{mele} E. J. Mele, \href{https://doi.org/10.1103/PhysRevB.84.235439}{Phys. Rev. B {\bf 84}, 235439 (2011)}.

\bibitem{Bistritzer2} R. Bistritzer and A. H. MacDonald, \href{https://doi.org/10.1103/PhysRevB.84.035440}{Phys. Rev. B {\bf 84}, 035440 (2011)}.

\bibitem{chu} Z.-D. Chu, W.-Y. He, and L. He, \href{https://doi.org/10.1103/PhysRevB.87.155419}{Phys. Rev. B {\bf 87}, 155419 (2013)}.

\bibitem{reina} G. Li, A. Luican, J. M. B. Lopes dos Santos, A. H. Castro Neto, A. Reina, J. Kong, and E. Y. Andrei, \href{https://doi.org/10.1038/nphys1463}{Nat. Phys. {\bf 6}, 109 (2010)}.

\bibitem{miller} D. L. Miller, K. D. Kubista, G. M. Rutter, M. Ruan, W. A. de Heer, M. Kindermann, P. N. First, and J. A. Stroscio, \href{https://doi.org/10.1038/nphys1736}{Nat. Phys. {\bf 6}, 811 (2010)}.

\bibitem{hermann} K. Hermann, \href{https://dx.doi.org/10.1088/0953-8984/24/31/314210}{J. Phys.: Condens. Matter {\bf 24}, 314210 (2012)}.

\bibitem{yan2} W. Yan, M. Liu, R.-F. Dou \textit{et al.}, \href{https://doi.org/10.1103/PhysRevLett.109.126801}{Phys. Rev. Lett. {\bf 109}, 126801 (2012)}.

\bibitem{hass} J. Hass, W. A. de Heer, and E. H. Conrad, \href{https://dx.doi.org/10.1088/0953-8984/20/32/323202}{J. Phys.: Condens. Matter {\bf 20}, 323202 (2008)}.

\bibitem{sprinkle}  M. Sprinkle \textit{et al.}, \href{https://doi.org/10.1103/PhysRevLett.103.226803}{Phys. Rev. Lett. {\bf 103}, 226803 (2009)}.

\bibitem{li2} X. Li, W. Cai, J. An \textit{et al.}, \href{https://doi.org/10.1126/science.1171245}{Science {\bf 324}, 1312 (2009)}.

\bibitem{lu2} C.-C. Lu, Y.-C. Lin, Z. Liu et al., \href{https://doi.org/10.1021/nn3059828}{ACS Nano {\bf 7}, 2587 (2013)}.

\bibitem{iwasaki} T. Iwasaki, A. A. Zakharov, T. Eelbo \textit{et al.}, \href{https://doi.org/10.1016/j.susc.2014.03.004.}{Surface Science {\bf 625}, 44 (2014)}.

\bibitem{sun} L. Sun, Z. Wang, Y. Wang \textit{et al.}, \href{https://doi.org/10.1038/s41467-021-22533-1}{Nat. Commun. {\bf 12}, 2391 (2021)}.

\bibitem{carozo} V. Carozo, C. M. Almeida, B. Fragneaud \textit{et al.}, \href{https://doi.org/10.1103/PhysRevB.88.085401}{Phys. Rev. B {\bf 88}, 085401 (2013)}.

\bibitem{trambly} G. T. de Laissardi\`{e}re, D. Mayou, and L. Magaud, \href{https://doi.org/10.1021/nl902948m}{Nano Lett. {\bf 10}, 804 (2010)}.

\bibitem{luican} A. Luican, G. Li, A. Reina \textit{et al.}, \href{https://doi.org/10.1103/PhysRevLett.106.126802}{Phys. Rev. Lett. {\bf 106}, 126802 (2011)}.

\bibitem{ash} G. Tarnopolsky, A. J. Kruchkov, and A. Vishwanath, \href{https://doi.org/10.1103/PhysRevLett.122.106405}{Phys. Rev. Lett. {\bf 122}, 106405 (2019)}.

\bibitem{dante} D. M. Kennes, J. Lischner, and C. Karrasch, \href{https://doi.org/10.1103/PhysRevB.98.241407}{Phys. Rev. B {\bf 98}, 241407(R) (2018)}.

\bibitem{cao2} Y. Cao, V. Fatemi, A. Demir \textit{et al.}, \href{https://doi.org/10.1038/nature26154}{Nature {\bf 556}, 80 (2018)}.

\bibitem{cao1} Y. Cao, V. Fatemi, S. Fang \textit{et al.}, \href{https://doi.org/10.1038/nature26160}{Nature {\bf 556}, 43 (2018)}.

\bibitem{Serlin} M. Serlin, C. L. Tschirhart, H. Polshyn \textit{et al.}, \href{https://doi.org/10.1126/science.aay5533}{Science {\bf 367}, 900 (2020)}.

\bibitem{shall2} S. Shallcross, S. Sharma, E. Kandelaki, and O. A. Pankratov, \href{https://doi.org/10.1103/PhysRevB.81.165105}{Phys. Rev. B {\bf 81}, 165105 (2010)}.

\bibitem{Ohta} T. Ohta, A. Bostwick, T. Seyller, K. Horn, and E. Rotenberg, \href{https://doi.org/10.1126/science.1130681}{Science {\bf 313}, 951 (2006)}.

\bibitem{zhang} Y. Zhang, T.-T. Tang, C. Girit et al., \href{https://doi.org/10.1038/nature08105}{Nature {\bf 459}, 820-823 (2009)}.


\bibitem{kuz} A. B. Kuzmenko, I. Crassee, D. van der Marel, P. Blake, and K. S. Novoselov, \href{https://doi.org/10.1103/PhysRevB.80.165406}{Phys. Rev. B {\bf 80}, 165406, (2009)}.

\bibitem{Castro} E. V. Castro, K. S. Novoselov, S. V. Morozov \textit {et al.}, \href{https://doi.org/10.1103/PhysRevLett.99.216802}{Phys. Rev. Lett. {\bf 99}, 216802 (2007)}. 

\bibitem{Mak} K. F. Mak, C. H. Lui, J. Shan, and T. F. Heinz, \href{https://doi.org/10.1103/PhysRevLett.102.256405}{Phys. Rev. Lett. {\bf 102}, 256405 (2009)}.

\bibitem{moon} P. Moon, Y.-W. Son, and M. Koshino, \href{https://doi.org/10.1103/PhysRevB.90.155427}{Phys. Rev. B {\bf 90}, 155427 (2014)}.

\bibitem{xian} L. Xian, S. Barraza-Lopez, and M. Y. Chou, \href{https://doi.org/10.1103/PhysRevB.84.075425}{Phys. Rev. B {\bf 84}, 075425 (2011)}.

\bibitem{San-Jose} P. San-Jose and E. Prada, \href{https://doi.org/10.1103/PhysRevB.88.121408}{Phys. Rev. B {\bf 88}, 121408(R) (2013)}.

\bibitem{Talkington} S. Talkington and J. Mele, \href{https://doi.org/10.1103/PhysRevB.107.L041408}{Phys. Rev. B {\bf 107}, L041408 (2023)}.

\bibitem{bhem2} G. Chen, L. Jiang, S. Wu et al., \href{https://doi.org/10.1038/s41567-018-0387-2}{Nat. Phys. {\bf 15}, 237 (2019)}.

\bibitem{bhem3} Y. Park, B. L. Chittari, and J. Jung, \href{https://doi.org/10.1103/PhysRevB.102.035411}{Phys. Rev. B {\bf 102}, 035411 (2020)}. 

\bibitem{bhem4} J. Shin, B. L. Chittari, and J. Jung, \href{https://doi.org/10.1103/PhysRevB.104.045413}{Phys. Rev. B {\bf 104}, 045413 (2021)}.


\bibitem{Chen} G. Chen, L. Jiang, S. Wu \textit{et al.}, \href{https://doi.org/10.1038/s41567-018-0387-2}{Nature Physics {\bf 15}, 237-241 (2019)}.

\bibitem{Chittari2} B. L. Chittari, G. Chen, Y. Zhang, F. Wang, and J. Jung, \href{https://doi.org/10.1103/PhysRevLett.122.016401}{Phys. Rev. Lett. {\bf 122}, 016401 (2019)}.

\bibitem{novo} K. S. Novoselov, A. K. Geim, S. V. Morosov \textit{et al.}, \href{https://doi.org/10.1038/nature04233}{Nature (London) {\bf 438}, 197 (2005)}.

\bibitem{McCann} E. McCann and V. I. Fal'ko, \href{https://doi.org/10.1103/PhysRevLett.96.086805}{Phys. Rev. Lett. {\bf 96}, 086805 (2006)}.

\bibitem{choi} M.-Y. Choi, Y.-H. Hyun, and Y. Kim, \href{https://doi.org/10.1103/PhysRevB.84.195437}{Phys. Rev. B {\bf 84}, 195437 (2011)}.

\bibitem{hass2} J. Hass, F. Varchon, J. E. Mill\'an-Otoya \textit{et al.}, \href{https://doi.org/10.1103/PhysRevLett.100.125504}{Phys. Rev. Lett. {\bf 100}, 125504 (2008)}.

\bibitem{gail} R. de Gail, M. O. Goerbig, F. Guinea, G. Montambaux, and A. H. Castro Neto, \href{https://doi.org/10.1103/PhysRevB.84.045436}{Phys. Rev. B {\bf 84}, 045436 (2011)}.

\bibitem{moon2} P. Moon and M. Koshino, \href{https://doi.org/10.1103/PhysRevB.85.195458}{Phys. Rev. B {\bf 85}, 195458 (2012)}.

\bibitem{lee} D. S. Lee, C. Riedl, T. Beringer \textit{et al.},  \href{https://doi.org/10.1103/PhysRevLett.107.216602}{Phys. Rev. Lett. {\bf 107}, 216602 (2011)}.

\bibitem{hejazi} K. Hejazi, C. Liu, and L. Balents, \href{https://doi.org/10.1103/PhysRevB.100.035115}{Phys. Rev. B {\bf 100}, 035115 (2019)}.

\bibitem{senthil} Y.-H. Zhang , H. C. Po, and T. Senthil, \href{https://doi.org/10.1103/PhysRevB.100.125104}{Phys. Rev. B {\bf 100}, 125104 (2019)}.

\bibitem{kubo1} R. Kubo, \href{https://doi.org/10.1143/JPSJ.12.570}{J. Phys. Soc. Jpn. {\bf 12}, 570 (1957)}.

\bibitem{kubo2} R. Kubo, M. Yokota, and S. Nakajima, \href{https://doi.org/10.1143/JPSJ.12.1203}{J. Phys. Soc. Jpn. {\bf 12}, 1203 (1957)}.


\bibitem{simao} S. M. Jo\~ao, M. An{\dj}elkovi\'c, L. Covaci, T. G. Rappoport, J. M. Viana Parente Lopes, and A. Ferreira, \href{https://doi.org/10.1098/rsos.191809}{R. Soc. Open Sci. {\bf 7}, 191809 (2020)}.

\bibitem{pires} J. P. Santos Pires, S. M. Jo\~ao, Aires Ferreira, B. Amorim, and J. M. Viana Parente Lopes, \href{https://doi.org/10.1103/PhysRevB.106.184201}{Phys. Rev. B {\bf 106}, 184201 (2022)}.

\bibitem{pires2} J. P. Santos Pires, S. M. Jo\~ao, Aires Ferreira, Amorim, and J. M. Viana Parente Lopes, \href{https://doi.org/10.1103/PhysRevLett.129.196601}{Phys. Rev. Lett. {\bf 129}, 196601 (2022)}.

\bibitem{zhu} X. Lin, H. Zhu, and J. Ni, \href{https://doi.org/10.1103/PhysRevB.101.155405}{Phys. Rev. B {\bf 101}, 155405 (2020)}.

\bibitem{shall} S. Shallcross, S. Sharma, and O. A. Pankratov, \href{https://doi.org/10.1103/PhysRevLett.101.056803}{Phys. Rev. Lett. {\bf 101}, 056803 (2008)}.

\bibitem{tomanek} X. Lin and D. Tom\'anek, \href{https://doi.org/10.1103/PhysRevB.98.081410}{Phys. Rev. B {\bf 98}, 081410(R) (2018)}.

\bibitem{rappoprt} J. H. Garc\'{\i}a, L. Covaci, and T. G. Rappoport, \href{https://doi.org/10.1103/PhysRevLett.114.116602}{Phys. Rev. Lett. {\bf 114}, 116602 (2015)}.

\bibitem{weisse} A. Wei\ss{}e, G. Wellein, A. Alvermann, and H. Fehske, \href{https://doi.org/10.1103/RevModPhys.78.275}{Rev. Mod. Phys. {\bf 78}, 275 (2006)}.

\bibitem{simao3} S. M. Jo$\tilde{a}$o and J. M. Viana Parente Lopes, \href{https://doi.org/10.1088/1361-648X/ab59ec}{J. Phys.: Condens. Matter {\bf 32}, 125901 (2020)}.

\bibitem{bastin} A. Bastin, C. Lewiner, O. Betbeder-Matibet, and P. Nozieres, \href{https://doi.org/10.1016/S0022-3697(71)80147-6}{J. Phys. Chem. Solids {\bf 32}, 1811 (1971)}.

\bibitem{ortmann} F. Ortmann, N. Leconte, and S. Roche, \href{https://doi.org/10.1103/PhysRevB.91.165117}{Phys. Rev. B {\bf 91}, 165117 (2015)}.

\bibitem{brihuega} I. Brihuega, P. Mallet, H. Gonz\'alez-Herrero \textit{et al.}, \href{https://doi.org/10.1103/PhysRevLett.109.196802}{Phys. Rev. Lett. {\bf 109}, 196802 (2012)}.

\bibitem{morell} E. Su\'arez Morell, P. Vargas, L. Chico, and L. Brey, \href{https://doi.org/10.1103/PhysRevB.84.195421}{Phys. Rev. B {\bf 84}, 195421 (2011)}.

\bibitem{peierls} R. Peierls, \href{https://doi.org/10.1007/BF01342591}{Z. Physik  {\bf 80}, 763 (1933)}.

\bibitem{ezawa} Z. F. Ezawa, \href{https://doi.org/10.1080/00107510903357796}{Quantum Hall Effects: Field Theoretical Approach And Related Topics, 2nd ed. (World Scientific, Singapore, 2008)}.

\bibitem{sinha} P. Sinha, S. Murakami, and S. Basu, \href{https://doi.org/10.1103/PhysRevB.102.085416}{Phys. Rev. B {\bf 102}, 085416 (2020)}.

\bibitem{do} T.-N. Do, P.-H. Shih, H. Lin, D. Huang, G. Gumbs, and T.-R. Chang, \href{https://doi.org/10.1103/PhysRevB.105.235418}{Phys. Rev. B {\bf 105}, 235418 (2022)}.

\end{thebibliography}
\end{document}